\documentclass[conference]{IEEEtran}
\usepackage{graphicx}
\usepackage{booktabs}
\usepackage{amsmath}
\usepackage{url}
\usepackage{listings}
\graphicspath{{figures/}}

\begin{document}

\title{Anticipating the Optimism Gap: Predicting Distribution-Shift Degradation
of RF-Impairment Detectors from In-Distribution Statistics}

\author{\IEEEauthorblockN{Chakshu Baweja}
\IEEEauthorblockA{Ashforde O\"U \\ contact@ashforde.org}}

\maketitle

\begin{abstract}
Detectors for GNSS radio-frequency impairments, including jamming, spoofing and
multipath, are usually reported with a single area under the ROC curve (AUC)
measured on the same distribution they were tuned on. That number tends to fall
once the operating conditions move, and the size of the drop is rarely known in
advance because labelled field data is scarce. We study this evaluation optimism
in a controlled setting and ask whether it can be predicted before any
out-of-distribution data is seen. On an open, parameter-grounded synthetic
testbed with a tunable severity shift, we evaluate thirteen detectors spanning
five physics baselines, full-feature logistic regression and multilayer
perceptrons, and single-feature learned controls, across four impairment classes.
Four findings emerge. The optimism gap, defined as the difference between the
in-distribution and shifted AUC, grows monotonically as the shift deepens (mean
Spearman correlation $0.50$ across detector and class pairs). The gap is driven by
how many observables a detector uses rather than by whether it is learned: a
single-feature learned detector and the physics baseline that reads the same
observable degrade by the same amount, while detectors that combine all five
observables degrade far less. The gap also varies systematically by impairment
class, with the deliberately hard matched-power spoofing case smallest. Finally,
and this is the central result, a simple ridge model built only from
in-distribution score statistics predicts the gap for a detector it has never seen
($R^2 = 0.47$) and for an impairment class it has never seen ($R^2 = 0.46$); both
are significant against a 2000-fold label-permutation null ($p < 0.001$) and both
survive removing the one feature that is, by construction, part of the quantity
being predicted. The headline findings are synthetic and we are explicit about that scope. We then
run the pre-registered protocol on three open field corpora. On the larger Jammertest
2024 campaign the cross-detector prediction holds on real data ($R^2 = 0.11$,
$p = 0.009$), while the cross-class prediction and a smaller, underpowered single-day
capture do not. On SatGrid, whose spoofer power sweep supplies the calibrated
severity axis the other field sets lack, in-distribution AUC overstates higher-severity
AUC across two independent recording sessions, by up to $0.22$ and to the point of sign
inversion at maximum spoofer power; across the eight pooled detector-by-session gaps the
in-distribution AUC and the realised gap are perfectly rank-correlated (Spearman
$\rho = 1.0$), the predicted direction, though four correlated detectors and a single
attack class leave the formal leave-one-detector-out predictor underpowered. So the mechanism survives contact with real data on the axis each
corpus can support, at a smaller magnitude than in simulation. We release the testbed,
a software-receiver front end that turns raw IQ into the correlator features the
signal-quality detector needs, the real-data ingest adapters, and the pre-registered
protocol.
\end{abstract}

\section{Introduction}
A practitioner choosing among GNSS interference detectors usually has one number to
go on: the AUC reported on a benchmark or a tuning set. It is widely understood
that this in-distribution figure flatters real performance, and recent work has
measured that discrepancy directly for learned GNSS interference
classifiers~\cite{discrepancies}. What a practitioner actually needs, and rarely
has, is a way to tell in advance how much a given detector's headline number will
deflate once conditions change. Field datasets that would answer the question are
expensive and often proprietary, so the gap remains anecdotal.

The machine learning community has built a body of methods that estimate a
classifier's out-of-distribution accuracy without out-of-distribution labels.
Examples include the linear in-distribution to out-of-distribution accuracy
correlation~\cite{aotl}, prediction from model agreement~\cite{agreement,gde},
thresholded-confidence estimators~\cite{atc}, and regression on dataset
statistics~\cite{denzheng,doc}. These methods target classification accuracy for a
fixed task and model evaluated across input distributions. We bring the same
question to GNSS impairment detection, where the natural figure of merit is AUC
rather than accuracy, and we change the transfer axis. Instead of predicting one
model's behaviour across distributions, we ask whether the gap can be predicted
for a detector, or an impairment class, that was held out of the prediction model
entirely. We borrow the controlled-corruption stance of ImageNet-C~\cite{imagenetc}
to isolate the shift variable cleanly, while being clear that our corpus is fully
synthetic and our claims are about mechanism rather than field magnitude.

This paper makes four contributions. We provide an open, reproducible,
parameter-grounded testbed for RF-impairment detectors under controlled
distribution shift, released as part of the open-source Kshana PNT-resilience
simulator~\cite{kshana}. We characterise how evaluation optimism grows with shift, what
governs it across detector families, and how it varies by impairment class, with
confidence intervals, a permutation null, and a matched-dimensionality control. We
show that the gap can be predicted from in-distribution statistics alone, and that
the prediction transfers to unseen detectors and unseen impairment classes, a
setting the accuracy-prediction literature does not address. We scope the work
honestly and supply a pre-registered protocol, naming concrete public datasets, so
that every claim is falsifiable on real data.

\section{Related work}
\textbf{Predicting performance under shift.}
A growing literature estimates out-of-distribution accuracy without labels.
Miller et al.~\cite{aotl} report a strong linear relationship between
in-distribution and out-of-distribution accuracy across many shifts. Baek et
al.~\cite{agreement} and Jiang et al.~\cite{gde} predict accuracy from the
agreement or disagreement of independently trained models. Garg et al.~\cite{atc}
threshold model confidence and prove identifiability limits. Deng and
Zheng~\cite{denzheng} regress accuracy on dataset-level feature statistics, and
Guillory et al.~\cite{doc} use a difference-of-confidences score. Calibration under
shift is studied by Ovadia et al.~\cite{ovadia}, and the broader empirics of
natural distribution shift by Recht et al.~\cite{recht}, Taori et
al.~\cite{taori}, and the WILDS benchmark~\cite{wilds}. This body of work fixes the
task and model and predicts across input distributions, and it predicts accuracy.
Our target is AUC, our features are detection-native (the margin between score
tails, the detection rate at a fixed false-alarm rate), and our transfer axis is
across detectors and across impairment classes. We do not claim priority in
predicting out-of-distribution performance from in-distribution statistics; our
claim is that this cross-detector and cross-class transfer for impairment-detection
AUC has not been shown before and is useful.

\textbf{GNSS interference detection.}
Jamming and spoofing detection is a mature field. Standard signatures include
carrier-to-noise drop, automatic-gain-control (AGC) excess, signal-quality
(correlator) distortion, and receiver-autonomous integrity residuals; see Kaplan
and Hegarty~\cite{kaplan} for an overview, Psiaki and Humphreys~\cite{psiaki} and
Dovis~\cite{dovis} for spoofing and countermeasures, Akos~\cite{akos} for
AGC-based detection, and recent surveys of learned methods~\cite{survey}. The
degradation of learned detectors on real data is documented
in~\cite{discrepancies}. Our task is distinct from out-of-distribution detection,
which flags individual anomalous inputs rather than forecasting a detector's
aggregate AUC under shift.

\textbf{ROC analysis.}
We use the Mann-Whitney form of AUC~\cite{hanley,bradley}; its variance follows
DeLong et al.~\cite{delong} with the fast computation of Sun and Xu~\cite{sunxu}.
Confidence intervals use the bootstrap~\cite{efron}, and our predictor is ridge
regression~\cite{ridge}.

\section{Testbed}
The corpus is class-balanced, labelled, and parameter-grounded. Each case is
described by five measurement-domain observables that a real receiver exposes:
jammer-to-signal ratio, effective carrier-to-noise drop, AGC power excess,
Early-minus-Late signal-quality imbalance, and a receiver-autonomous integrity
parity statistic. These are produced by composing the jamming, AGC, signal-quality
and parity models of Kshana, an open-source simulator for positioning, navigation
and timing (PNT) resilience~\cite{kshana}, with seeded Gaussian measurement noise.
No raw IQ or field captures are used. There are five classes: nominal, jamming, time
spoofing (matched power, the deliberately hard near-nominal case), position
spoofing, and multipath. A scalar severity scale in $(0,1]$ multiplies every
impaired-class severity. At a scale of $1.0$ the corpus is in-distribution and
reproduces bit-for-bit; smaller scales produce subtler impairments.

The panel contains thirteen detectors. Five are transparent physics baselines:
carrier-to-noise energy, AGC excess, signal-quality imbalance, integrity parity,
and a documented maximum-of-standardised-statistics combiner. Five are full-feature
learned detectors: logistic regression trained by full-batch gradient descent, and
four multilayer perceptrons at hidden widths of $4$, $8$, $16$ and $32$. Three are
single-feature learned controls: logistic regression restricted to one observable
each, at matched input dimensionality to the single-observable physics baselines.
Learned detectors are trained on an in-distribution training split only.

Every AUC reported here is computed over model-derived labels on synthetic data.
The corpus is a separability and pipeline harness rather than a difficulty
benchmark, so a high in-distribution AUC shows that a detector reads the right
observable, not that it would perform in the field. Two leakage guards protect the
train and test partition: an exact-key guard and a near-duplicate observable guard
using a per-component infinity norm, restricted within a class. Both are asserted
on every training split.

\section{Methods}
For a detector $d$, class $c$ and severity $s$, let $\mathrm{AUC}_c(d,s)$ be the
Mann-Whitney AUC of $d$'s scores on class-$c$ positives against nominal negatives
drawn at severity $s$. The in-distribution value $\mathrm{AUC}_c(d,1.0)$ is always
computed on a held-out test split. The optimism gap is
\begin{equation}
\Delta(d,c,s) = \mathrm{AUC}_c(d,1.0) - \mathrm{AUC}_c(d,s), \quad s < 1.
\end{equation}

We test four hypotheses. H1: $\Delta$ increases with the shift magnitude $1-s$.
H2: detector evidence breadth, not the use of learning, governs $\Delta$. H3:
$\Delta$ differs systematically by class. H4: an in-distribution-only regressor
predicts $\Delta$ for held-out detectors and held-out classes.

The experiment grid crosses $13$ detectors, $4$ classes, severities in
$\{0.2,0.4,0.6,0.8\}$, and $5$ seeds, with $400$ cases per class and a $0.7$ train
fraction. For each seed we generate the in-distribution corpus, form a stratified
split (asserting the leakage guard), train the learned detectors, measure held-out
in-distribution per-class AUC, and then for each severity generate an independent
out-of-distribution corpus and record the gap. We aggregate per cell as the mean
gap with a percentile bootstrap interval and an across-seed standard error, and
per detector-class pair we fit a trend (Spearman correlation, ordinary
least-squares slope, and the mean in-distribution AUC).

Every statistical routine is checked against a closed form: the binormal identity
$\mathrm{AUC}=\Phi(d'/\sqrt2)$ for the Mann-Whitney estimator, a hand-worked DeLong
variance, a textbook tied-rank Spearman value, and exact ordinary least-squares
recovery for the ridge solver. No third-party statistics dependency is used.

\textbf{The gap predictor.}
For each detector-class pair we extract six in-distribution-only features from the
held-out test set: the in-distribution AUC; the separation
$d'=(\mu_{+}-\mu_{-})/\sigma_{\text{pool}}$; the overlap between the positive and
negative score distributions; their variance ratio; the tail margin
$(q_{05}^{+}-q_{95}^{-})/\sigma_{\text{pool}}$; and the detection rate at a target
false-alarm rate. A seventh, ablatable feature is a self-perturbation slope,
estimated from a mild self-imposed micro-shift using independently seeded probe
corpora and no out-of-distribution labels. The target is the realised gap averaged
over the severity sweep. A ridge regressor predicts it, with features standardised
on the training fold only and the penalty fixed a priori at $0.1$ rather than tuned
on test folds. We validate with leave-one-detector-out and leave-one-class-out
cross-validation against a predict-the-mean baseline, and we assess significance
with a 2000-fold label-permutation null. Because the target contains the
in-distribution AUC as one additive term, we also report a shape-only predictor
that omits it; its survival shows the predictability is not a definitional artefact.

\section{Results}
All values come from the released artifact, which records $208$ cells, $52$ trends
and $260$ predictor rows over five seeds. Figures are regenerated from that artifact.

\begin{figure}[t]\centering\includegraphics[width=\columnwidth]{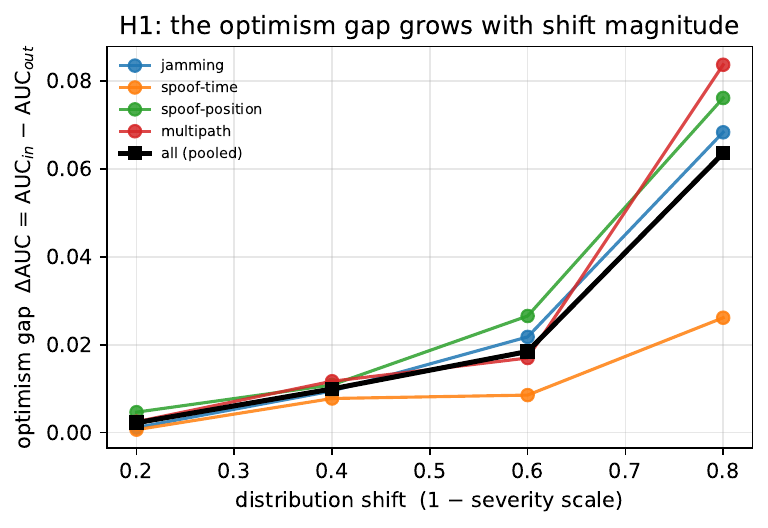}
\caption{The optimism gap grows as the distribution shift deepens, shown per class
and pooled.}\label{fig:scaling}\end{figure}

\textbf{H1: the gap grows with shift (Fig.~\ref{fig:scaling}).}
The pooled mean gap rises from $0.002$ at severity $0.8$ to $0.010$, $0.019$ and
$0.064$ at $0.6$, $0.4$ and $0.2$. Across the $52$ detector-class trends the
Spearman correlation between $1-s$ and $\Delta$ has mean $0.50$, with $42$ of $52$
positive and $30$ significant at the five-percent level. We describe this as a
trend rather than a law because the few non-positive cases coincide with a
near-chance in-distribution AUC, that is, detectors with no separability to lose;
the artifact records each trend's mean in-distribution AUC so this is checkable.
Monotonicity on its own is unsurprising, since attenuating the observable a
detector reads must lower its AUC. The informative part, and what makes the
predictor in H4 possible, is that the rate of collapse varies widely across
detectors at a comparable in-distribution AUC.

\begin{figure}[t]\centering\includegraphics[width=\columnwidth]{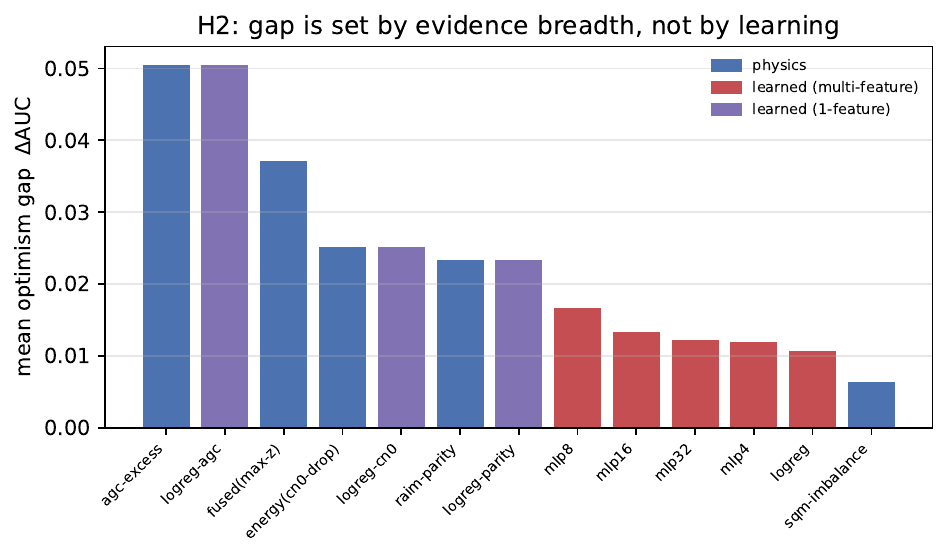}
\caption{Per-detector gap by family. The matched physics and one-feature-learned
pairs coincide, and the full-feature learned detectors degrade least.}\label{fig:family}\end{figure}

\textbf{H2: evidence breadth, not learning (Fig.~\ref{fig:family}).}
Pooled over classes and severities, the mean gap is $0.029$ for the physics
baselines, $0.033$ for the single-feature learned controls, and $0.013$ for the
full-feature learned detectors. The clearest evidence is the matched-dimensionality
control in Table~\ref{tab:matched}: a single-feature learned detector and the
physics baseline that reads the same observable have the same gap to three decimal
places. This is expected, since a one-feature logistic detector is a monotone
transform of that observable and therefore has the same per-class AUC, and it makes
the point precisely. What sets the gap is how much evidence a detector integrates,
not whether it was trained. Narrow detectors, learned or not, are fragile; the
detectors that combine all five observables retain far more of their separability
under shift. This refines the common intuition rather than contradicting it.
Trained models on real data can additionally overfit nuisance correlations that a
controlled generator does not contain, an outcome our protocol in
Section~\ref{sec:limits} pre-registers.

\begin{table}[t]\centering
\caption{Matched-dimensionality control. At equal input dimensionality a learned
detector and a physics baseline reading the same observable have the same gap.}
\label{tab:matched}
\begin{tabular}{lcc}\toprule
observable & physics gap & one-feature learned gap \\\midrule
carrier-to-noise drop & 0.025 & 0.025 \\
AGC excess & 0.051 & 0.051 \\
integrity parity & 0.023 & 0.023 \\\bottomrule
\end{tabular}
\end{table}

\begin{figure}[t]\centering\includegraphics[width=0.86\columnwidth]{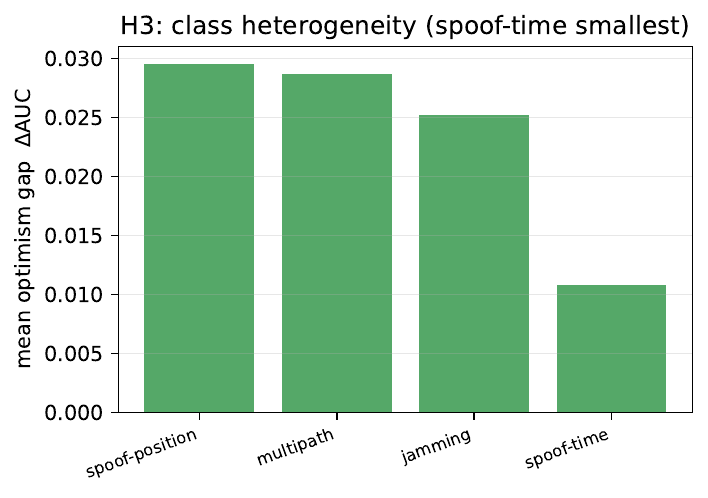}
\caption{Per-class mean gap. Time spoofing is smallest.}\label{fig:perclass}\end{figure}

\textbf{H3: class heterogeneity (Fig.~\ref{fig:perclass}).}
The mean gap is $0.030$ for position spoofing, $0.029$ for multipath, $0.025$ for
jamming, and $0.011$ for time spoofing. The matched-power time-spoofing case, which
is the hardest to separate even in-distribution, has the least optimism to lose.
The classes with the most impressive in-distribution numbers are the ones that fall
furthest, which is a useful caution for anyone ranking detectors on
in-distribution AUC alone.

\begin{figure*}[t]\centering\includegraphics[width=0.86\textwidth]{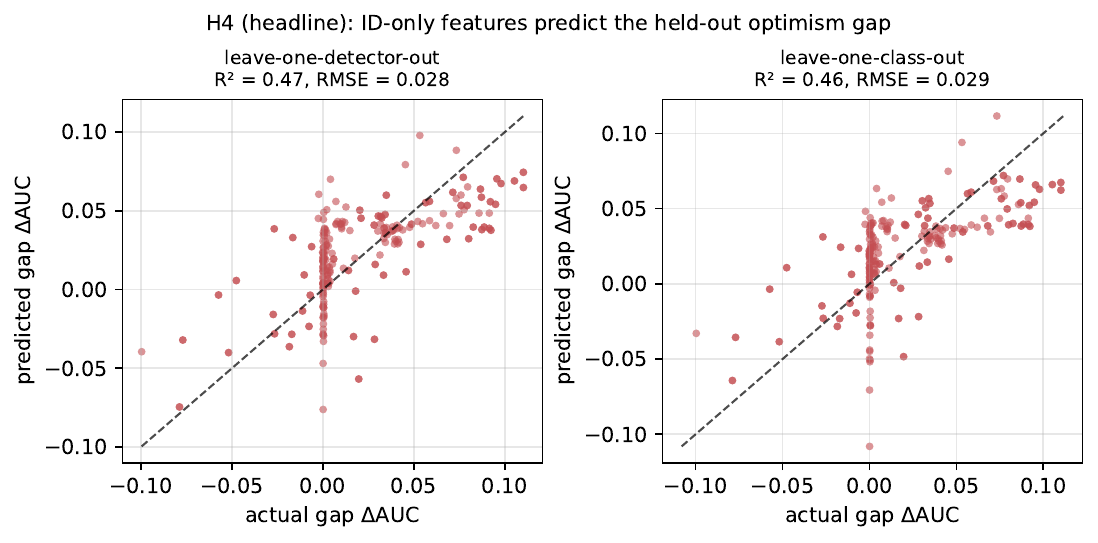}
\caption{Predicted against actual gap for both cross-validation splits. Each point
is a held-out detector-class-seed row; the dashed line is parity.}\label{fig:predictor}\end{figure*}

\textbf{H4: predicting the gap (Fig.~\ref{fig:predictor}).}
A ridge model built only from in-distribution statistics predicts the held-out gap
for a detector and for a class that were excluded from training.
Leave-one-detector-out gives $R^2 = 0.47$ at a root-mean-square error of $0.028$
over $13$ folds; leave-one-class-out gives $R^2 = 0.46$ at $0.029$ over $4$ folds.
Both beat the predict-the-mean baseline, and across $2000$ label permutations none
reached the observed $R^2$, so $p \approx 5\times10^{-4}$ in each case. The
cross-detector result is much stronger than in a seven-detector pilot, where it was
$0.26$, as expected when the number of folds grows.

The result is not an artefact of the in-distribution AUC feature. Since the gap is
the in-distribution AUC minus the shifted AUC, that feature is one additive term of
the target, and it carries the largest standardised coefficient at $0.13$. Refitting
without it, the shape-only predictor still reaches $R^2 = 0.34$ for detectors and
$0.27$ for classes, so genuine predictability lives in the shape of the score
distribution. Removing instead the one feature that touches the generator, the
self-perturbation slope, barely changes the full result, from $0.47$ to $0.45$ and
$0.46$ to $0.43$, so the prediction does not depend on probing the simulator. Among
the shape features the score overlap and the tail margin carry the most weight, with
standardised coefficients of $+0.07$ and $-0.03$. In plain terms, a high
in-distribution AUC achieved on a thin, heavily overlapping score margin is the
signature of a detector that will degrade, and that signature is visible before any
shifted data is collected.

\section{Limitations and falsification protocol}\label{sec:limits}
The corpus is synthetic from end to end. Unlike ImageNet-C, which corrupts real
images, every observable here is generated, so our claims concern mechanism and
regime, not the magnitude of any real field gap. The shift is a controlled severity
change, which makes the gap synthetic-to-synthetic. Replication uses five seeds and
four severity levels, so the per-cell intervals are wide and should be read as a
five-seed spread; the cross-class result rests on four folds. The single-feature
learned controls equal the physics baselines by construction, so that control
demonstrates the mechanism rather than measuring it independently.

A faithful real-data test needs a labelled corpus in the same measurement domain,
with a defined shift axis. To make that test runnable rather than hypothetical, we
built the receiver-domain ingest stage and released it with the simulator. A set of
adapters reads the public file formats, RINEX observation and broadcast navigation,
u-blox UBX receiver logs, Android GnssLogger captures, and software-receiver
correlator dumps, and maps each available channel to the observables used here:
carrier-to-noise, AGC, the u-blox jamming indicator, the Early-minus-Late
signal-quality metric, and a pseudorange RAIM statistic. Each channel reuses the
same engine the synthetic corpus uses, so a real observation passes through the
identical scoring path. The pseudorange RAIM channel is exercised on a surveyed IGS
station, where it solves the broadcast-ephemeris position and returns a per-epoch
consistency statistic, which fixes the one channel that needs a full position solve
rather than a single measurement. Because no single public set exposes all five
observables, the ingest scores each available channel independently, which matches
the ragged coverage of the field record: carrier-to-noise is widely logged, AGC
appears only on receiver logs, and the signal-quality metric needs tracked
correlators. We ran this probe on three open, labelled field corpora. The first is the Yunnan
University spoofing and jamming capture~\cite{yunnan}, a single rooftop u-blox
receiver recorded across a day of staged attacks, from which we take carrier-to-noise
per constellation and band as the detector panel and split severity by the documented
attack phases. The second is the Jammertest 2024 campaign~\cite{jammertest}, a
national over-the-air test with four attack classes (jamming, spoofing, meaconing, and a combined attack)
recorded both stationary and dynamic, from which we take carrier-to-noise, AGC, and
the jamming indicator, and use the change from a stationary to a dynamic receiver as
the distribution shift. For these two the clean negatives are the receiver's own
pre-attack and inter-attack seconds, so each comparison is like for like. The third
is SatGrid~\cite{satgrid}, which records genuine GPS L1 alongside a spoofer replayed
at several amplification levels across independent recording sessions; its per-channel
software-receiver tracking dumps give carrier-to-noise, the Early-minus-Late
signal-quality metric, the carrier-lock test, and a prompt quadrature ratio, with the
genuine recording as the negatives and the amplification level as the calibrated
severity axis the other two corpora lack. We use its two Arlington sessions and pool
the per-detector gaps across them. To
score the signal-quality metric from raw antenna samples, where a dataset ships them
rather than tracking dumps, we built and released a software-receiver front end, C/A
code generation, code-phase by Doppler acquisition, and a closed delay-lock and Costas
tracking loop, that turns raw IQ into the Early/Prompt/Late correlator taps the metric
reads. It is validated on synthetic IF, where a clean signal scores a near-zero
imbalance and an injected multipath echo raises it, and it consumes the SatGrid
tracking dumps directly. The TEXBAT and OAKBAT raw-signal batteries remain the route
to running that front end on tens of gigabytes of field IQ; we verified their access
and format but did not fetch them here. The probes below were pre-registered before
any corpus was scored.

We pre-register the following tests. Fix the shift axis, the sample sizes and the
primary endpoint before unblinding. H1 is refuted if fewer than half of the trends
with in-distribution AUC above $0.6$ are positive, or the mean correlation is not
positive. H2 is refuted if, on real data, full-feature learned detectors show a
larger gap than narrow detectors at matched in-distribution AUC. H3 is refuted if
the class ordering does not rank-correlate positively with ours. The primary
endpoint H4 is refuted if either leave-one-out $R^2$ fails to beat predict-the-mean
at permutation $p<0.05$. The in-repo artifacts carry machine-checked status labels,
and the no-overclaim and demonstrator continuous-integration checks remain green.

\begin{figure}[t]\centering\includegraphics[width=0.86\columnwidth]{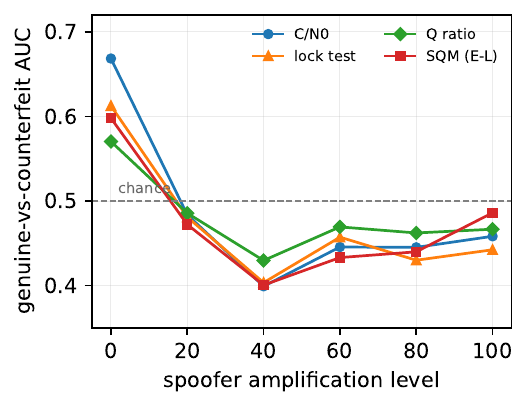}
\caption{Graded optimism on real data (the SatGrid Arlington session that spans matched
to maximum spoofer power). Each tracking-channel detector separates genuine from
counterfeit at the lowest spoofer power (level $0$, AUC $0.57$ to $0.67$) but falls
below chance as the spoofer is amplified, because a strong spoof looks healthier than
the authentic signal. Calibrating a detector at level $0$ and deploying it at higher
levels therefore overstates its AUC, by $0.16$ on average. The dashed line is
chance.}\label{fig:satgrid}\end{figure}

Read against these criteria, the real data give a partial and honest confirmation.
On Jammertest 2024, the richer corpus with fifty detector-by-class gap samples over
thirteen detectors and four attack classes, the in-distribution statistics predict
the leave-one-detector-out gap at $R^2 = 0.11$ with permutation $p = 0.009$, so H4
holds on the detector axis. The same predictor does not transfer across the four
classes ($R^2 = -0.25$), which differ too much in mechanism for one to forecast
another from four folds, so H4 is refuted on the class axis. The Yunnan capture,
with twenty samples and a coarser severity proxy, is underpowered: its
leave-one-detector-out $R^2$ is $-0.25$ at $p = 0.19$, neither confirming nor
cleanly refuting. SatGrid adds the graded-severity reading the other two cannot:
calibrating each detector at the lowest spoofer power in a session and deploying it at
higher powers, the in-distribution AUC overstates the realized AUC. In the session that
spans matched to maximum power, the gap reaches $0.16$ on average and $0.22$ at most,
and at the strongest amplification the realized AUC falls below $0.5$ as the strong
spoof looks healthier than the genuine signal (Fig.~\ref{fig:satgrid}). The second
session is confined to high power, where detection has already failed, so its gaps are
near zero, exactly because its in-distribution reference sits in the already-failed
regime, which the in-distribution AUC itself registers. Pooling the eight
detector-by-session gaps, the in-distribution AUC and the realized gap are perfectly
rank-correlated (Spearman $\rho = 1.0$), the direction H4 predicts, though with four
correlated detectors and a single attack class the leave-one-detector-out ridge still
cannot beat predict-the-mean ($R^2 = -0.13$, $p = 0.10$), so SatGrid confirms the
graded gap and its direction across sessions, not the quantitative predictor. The detector-domain physics is sound throughout: meaconing
carrier-to-noise separates at AUC near $0.94$, AGC is a strong discriminator, and a
GPS-only spoofer leaves Galileo carrier-to-noise near chance, the expected control.
The real cross-detector effect is genuine but smaller than the synthetic $R^2=0.47$,
as expected when the shift is an uncontrolled field condition (receiver mobility, or
a documented attack phase) rather than a calibrated severity sweep. We read this as
the mechanism surviving contact with real data on the axis the data can support, and
as a caution against reading the synthetic magnitude as a field number.

\section{Reproducibility}
Every number regenerates from a single command, and the figures are built from the
resulting artifact:
\begin{lstlisting}
cargo run --release --example optimism_study
python3 make_figures.py
\end{lstlisting}
The artifact is versioned, self-describing JSON. It records all aggregates, the
trends with their mean in-distribution AUC, the family gaps and the
matched-dimensionality pairs, the predictor's cross-validation with the full
predicted-against-actual scatter, the permutation-null p-values, the shape-only and
no-self-slope ablations, intercept-labelled coefficients, and provenance covering
the engine version, a configuration hash, and the seeds. The statistical routines
are unit-tested against closed forms, and the corpus is reproducible bit-for-bit
from its configuration and seed. The testbed, including the corpus generator, the
detector panel, the statistics and the gap predictor, is released under AGPL-3.0 as
part of Kshana~\cite{kshana}.

The real-data probes follow the same pattern. Each public corpus has a dedicated
ingest example that reuses the same scoring path, so a result regenerates from the
downloaded files in one command:
\begin{lstlisting}
cargo run --release --example yunnan_probe \
    -- windows.json out.json observation12.json \
       observation16.json observation17.json
cargo run --release --example jammertest_probe \
    -- <dataset_root> out.json stationary
python3 satgrid_extract.py feats.csv \
    <session_dir> [<session_dir> ...]
cargo run --release --example satgrid_probe \
    -- feats.csv out.json
cargo run --release --example texbat_probe -- \
    out.json 25000000 8 0 \
    nominal:cleanStatic.bin p10:ds2.bin \
    p1_3:ds3.bin p0_4:ds4.bin
\end{lstlisting}
A generic \texttt{ingest\_realdata} manifest path is also provided for arbitrary
file sets. The adapters carry their own unit tests with hand-derived oracles, the
RAIM channel is checked against a surveyed IGS station, and the software-receiver
front end is checked end to end on synthetic IF (acquire, track, and a known
clean-versus-multipath signal-quality contrast), so the path from a public file or
raw sample to a scored observation is testable independently of the analysis. For
SatGrid the only non-Rust step is a small HDF5-to-CSV extractor, so the published
crate carries no HDF5 dependency. The AUC is computed in rank form, so a real-data
group of hundreds of thousands of samples scores in the same pass as a synthetic cell.

\section{Conclusion}
Evaluation optimism in RF-impairment detection is usually treated as a caveat. We
show it is measurable and, more usefully, predictable. From in-distribution
statistics alone a simple model forecasts how far a detector's AUC will fall under a
controlled shift, and the forecast transfers to detectors and impairment classes it
never saw during fitting, at $R^2$ near $0.47$ and $0.46$ with permutation
$p<0.001$, surviving the removal of the feature that is part of the target by
construction. A matched-dimensionality control shows the gap is set by how much
evidence a detector integrates rather than by whether it is trained. The study is
synthetic and we say so plainly, and we ship a pre-registered protocol that names
concrete public datasets so the community can confirm or refute it. The practical
message is simple: a high in-distribution AUC bought on a thin, overlapping score
margin should be read as a warning, and it can be read before deployment.

\end{document}